\documentclass[twocolumn,aps,pre,showpacs,preprintnumbers,floatfix,superscriptaddress]{revtex4-1}

\usepackage{amsmath}
\usepackage{amssymb}
\usepackage{graphicx}
			
\usepackage{xspace}
\newcommand{\eg}{{e.g.,\/}\xspace}
\newcommand{\ie}{{i.e.,\/}\xspace}
\newcommand{\etal}{{\it et~al.\/}\xspace}

\newcommand{\Eq}[1]{Eq.~(\ref{#1})}

\newcommand{\Fig}[1]{Fig.~\ref{#1}}
\newcommand{\Sec}[1]{Sec.~\ref{#1}}

\begin{document}

\title{Reducing parametric backscattering by polarization rotation}

\author{Ido Barth}
\email{ibarth@princeton.edu} 

\author{Nathaniel~J. Fisch}

\affiliation{Department of Astrophysical Sciences, Princeton University, Princeton, New Jersey 08540, USA}

\date{\today}

\begin{abstract}
When a laser passes through underdense plasmas, Raman and Brillouin Backscattering can reflect a substantial portion of the incident laser energy.
This is a major loss mechanism, for example, in employing lasers in inertial confinement fusion.
However, by slow rotation of the incident linear polarization, the overall reflectivity can be reduced significantly.
Particle in cell simulations show that, for parameters similar to those of indirect drive fusion experiments, polarization rotation reduces the reflectivity by a factor of $5$.
A general, fluid-model based, analytical estimation for the reflectivity reduction agrees with simulations.
However, in identifying the source of the backscatter reduction, it is difficult to disentangle the rotating polarization from the frequency separation based approach used to engineer the beam's polarization.
Although the backscatter reduction arises similarly to other approaches that employ frequency separation, in the case here, the intensity remains constant in time.

\end{abstract}


\maketitle

\section{Introduction}
Mitigating parametric backscattering (PB) is one of the major challenges in light-matter interaction.
For example, in inertial confinement fusion (ICF), PB is responsible for reflecting of about $10\%$ of the total incident lasers energy \cite{Hinkel_PoP_2011,Kirkwood_PPCF_13,Hinkel_PPCF_13}.
As a result, less energy is available for the hydrodynamic pellet compression.
There are two main types of PB, stimulated Raman backscattering (SRS) and stimulated Brillouin backscattering (SBS). 
In SRS the incoming energy is backscattered through an electron plasma wave (EPW) while the energy transfer mediator in SBS is an ion acoustic wave (IAW).
Because of the large ion/electron mass ratio, SRS generally has a higher growth rate than SBS, and therefore generally dominant. 
In case that SRS is Landau damped (for high electron temperatures) or excluded (for over quarter-critical densities), SBS can become the dominant scattering mechanism.
In the parameter regime relevant to the current design of indirect ICF, SRS is dominant for the inner beams while SBS is dominant for the outer beams \cite{Hinkel_PoP_2008,Hinkel_PoP_2011}.
When counter propagating beams cross, PB depends on the relative polarization between the beams \cite{Kirkwood_PoP_2006}.
Multibeam SRS, which was recently recognized as a possible scattering mechanism in indirect-drive ICF, also depends on the polarization arrangement of the crossing beams \cite{Michel_PRL_2015}.

PB, as other laser plasma interactions, can be controlled by manipulating  the laser and/or the plasma. 
On the plasma side, saturation of the plasma wave limits the overall reflectivity, where possible saturation of PB mechanisms include nonlinear Landau damping, collisions, and the Langmuir decay instability \cite{Kirkwood_PoP_2003}.
For example, introducing a thin layer of borated gold on the inside of the hohlraum wall increases the electron temperature and thus enhances the Landau damping of the EPW and reduces SRS \cite{Hinkel_PPCF_13}.
On the laser side, multiple manipulation techniques of the incident laser are used to mitigate PB including 
 smoothing methods as spatial smoothing \cite{Dixit_OL_1994}, spectral dispersion \cite{Skupsky}, polarization smoothing \cite{Lefebvre,Fuchs}, and a combination of these three smoothing methods \cite{Berger_PoP_1999,Moody_PRL_2001}.
Interestingly, perpendicular polarizations can also mitigate cross beam energy transfer, and thereby affect symmetry \cite{Michel_PRL_2014}.
Other methods to reduce PB comprise modulation of the laser intensity such that the effective interaction length is reduced, thereby, diminishing the effective growth rate and the reflectivity.
An example of such a method is the spike trains of uneven duration and delay (STUD) pulses, in which, the laser energy is delivered via a train of short but more intense pulses separated by off-periods  \cite{STUD1,STUD2,STUD3}.
In all of the aforementioned methods the linear polarization of the beam is kept constant.

In this paper, we take advantage of the two possible polarizations in order to reduce PB by reducing the effective interaction length, while keeping the pulse intensity constant.
This is done by slowly rotating the linear polarization of the incident laser. 
The idea is that, on the one hand, only the polarization component that is parallel to the incident light polarization is getting amplified (via SBS or SRS), while, on the other hand, light cannot change its polarization when traveling through a non-magnetized plasma.
The combination of these two facts implies that slowly rotating or oscillating the incident beam polarization  reduces the effective interaction length between the incident laser and a noise component.
As a result, the parametric instability noise amplification is depressed, resulting in a reduction in the overall reflectivity.
For significant reflectivity reduction, the rotation frequency should be of the order of the parametric growth rate or larger.
Although, for rotating polarization, both transverse polarizations contribute to the noise amplification, the exponential character of PB results in a reduction of the overall reflectivity.
In addition, because the effect is accumulative in the growth rate itself, the reduction becomes more prominent for systems with less initial noise but more e-foldings.
The main advantage of the proposed method over other methods to reduce PB, \eg smoothing by spectral dispersion \cite{Skupsky}, is that significant reduction can be achieved without requiring large bandwidth  that is be beyond current technology.  

The paper is organized as follows.
\Sec{sec2} introduces the rotating polarization light.
\Sec{sec3} studies the effect on PB  via one-dimension (1D) particle-in-cell (PIC) simulations and discusses the relevant parameter regimes.
A theoretical approach is developed in \Sec{sec4} based on the 3-wave interaction fluid model in order to estimate the reduction in the overall reflectivity. 
\Sec{sec5} introduces an alternative, spectral-based approach while \Sec{sec6} presents additional results and discusses other regimes.
\Sec{sec7} summarizes the conclusions.


\section{Rotating polarization light} \label{sec2}
Consider a vector potential, $\mathbf{A}$, of form
\begin{equation} \label{A0}
	\mathbf{A}_0 = \Re \left[ a_0 \, \mathbf{p} \,  e^{i(\omega_0 t - k_0 x)} \right],
\end{equation}
where, $\Re$ denotes the real part; $\mathbf{A}_0$ is in the units of $m_e c^2 /e$; 
$m_e$ and $e$ are the electron mass and charge, respectively;
 $c$ is the speed of light; 
$a_0$ is the (complex) dimensionless amplitude;
$\omega_0$ and $k_0$ are the laser frequency and wave number, respectively;
and $\mathbf{p}$ is the (time dependent) polarization unit vector,
\begin{equation} \label{p}
    \mathbf{p} = \left( \hat{y}\cos\Theta + \hat {z}\sin\Theta \right).
\end{equation}
Here, $\hat{y}, \hat{z}$ are unit vectors in the transverse direction and $\Theta=\Theta(t,x)$ is the polarization rotation angle. 
For simplicity, we consider a simple polarization rotation such that the polarization angle at the plasma boundary is given by $\Theta(t,x=0)=\Omega \, t$, where  $\Omega$ is the rotating frequency.
In other words, in the vacuum (outside the plasma) the polarization angle can be written as $\Theta=\Omega t'$, where $t'=t-x/c$.
It is important to note that since $\Omega \ne \omega_0$, the light polarization is not a circular.
Moreover, for $\Omega\ll\omega_0$ the light can be considered as locally linearly polarized, while its polarization rotates at frequency $\Omega$.

Notably, such type of light can be realized by combining two counter-rotating, circularly polarized lasers with frequency difference of $2\Omega$,
\begin{equation} \label{A_pm}
    A_\pm=\frac{a_0}{2} \left[\hat{y} \cos(\omega_\pm t') \pm \hat{z} \sin(\omega_\pm t') \right],
\end{equation}
where $\omega_\pm = \omega_0 \pm \Omega$. This is because
\begin{equation} \label{A_pm_a}
    A_{+}+A_{-} = a_0 \cos(\omega_0 t') \left[\hat{y} \cos(\Omega t') + \hat{z} \sin(\Omega t') \right] = \mathbf{A}_0.
\end{equation}
Eqs.~({\ref{A_pm})-(\ref{A_pm_a}) suggest a practical way to create such a rotatingly polarized pulse.
First, the incident, linearly polarized, pulse is split into two perpendicular linear polarizations.
Second, the beams are frequency shifted, one up and one down.
Then, two wave plates are used to convert the beams into right and left circularly polarized.
Finally, the two beams, $A_\pm$, are recombined into a single beam, $A_0$, with rotating polarization.
Similar technique but with additional frequency chirp is employed in molecular optical centrifuges \cite{optical_centrifuge}.

Alternative realizations of such an optical rotation employ, for example, Faraday rotation with time-dependent magnetic field or laser frequency chirp, chirping the pulse frequency and passing through chiral materials, or some combination of these techniques. 
However the latter techniques might be more challenging because of the short rotation time scale that is required to reduce PB.

Note that by replacing the sum of two circularly polarized in \Eq{A_pm_a} with the sum of two linearly polarized waves of the same intensity one gets a beat wave pulse, \ie  $a_0 \hat{y} \cos(\omega_{+} t')+a_0 \hat{y} \cos(\omega_{-} t') = 2 a_0 \cos(\omega_0 t') \hat{y} \cos(\Omega t') $, where the envelop amplitude of the beat wave is as twice higher for the same total energy.
This type of modulation is similar to the STUD pulses in the sense that the laser intensity is modulated in time 
having approximately half time on and half time off but with sinusoidal envelop instead of flattop square envelope. 
This amplitude modulation reduces the effective PB interaction length.
Therefore the beat wave is anticipated to result in a reduction in the reflectivity.
However, in this paper we focus on pulses of constant intensities and leave the detailed comparison between different methods to a future study.

\section{PIC simulations} \label{sec3}

To illustrate the PB reduction effect, we consider parameters similar to those of ICF experiments and compare incident pulse of constant polarization (Fig.~\ref{fig1}) with pulse with rotating polarization (\Fig{fig2}). 
The 1D PIC simulations were run with the code EPOCH \cite{epoch} with $120$ cells per $\mu$m and $40$ particles per cell.
In addition to the physical noise, PIC simulations introduce also numerical noise, which depends on the resolution. 
Therefore, to preserve the initial background noise, we use the same resolution in all simulations and compared the reduction in the reflectivity.

The physical parameters in the simulations were as follows.
The laser wavelength and intensity were $\lambda_0=0.351 \, \mu$m and $I_0=5\times10^{14}$~ W/cm$^2$;
the pulse duration was $\tau=15$~ps;
the plasma  length and density were $L=0.5$~mm and $n_0=5\times10^{20} \,{\rm cm}^{-3}$;
the electron (ion) temperature was $T_e=1000$ eV  ($T_i=50$ eV).
In \Fig{fig1}, we present a non-rotating pulse ($\Omega=0$) just after passing the plasma. 
In the figure, the plasma boundaries are depicted by dashed lines and the pulse propagates from left to right.
Thus, the left part of \Fig{fig1}a represents the reflected light while the right part is the transmitted pulse.
Spectral analysis of the reflected part that is shown in \Fig{fig1}b suggests that the dominant backscattering mechanism is this case is SBS. 
This is because, in this regime, SRS is Landau damped due to the high electron temperature.

We define the reflectivity 
\begin{equation}
    R=\frac{U_{\rm r}}{U_{\rm i}},
\end{equation}
where, $U_{\rm i}$ and $U_{\rm r}$ are the incident and reflected electromagnetic fluences, respectively. 
These fluences are calculated by integrating the incident and reflected intensities over time.
For example, the reflectivity in the case presented in \Fig{fig1} is $R=0.15$.

Now, we rotate the incident beam with frequency $\Omega/2 \pi= 500$GHz and present the electric fields of the two linear polarizations, $\hat{y}$ and $\hat{z}$ in \Fig{fig2}a. 
The overall reflectivity is reduced to the value of $R=0.03$, illustrating the effect of reflectivity reduction. 

We anticipate also a $\Omega$-dependency of the reflectivity. 
For rotation that is too slow, the polarization is effectively fixed for many e-foldings.
Therefore, it is clear that in the limit $\Omega\ll\gamma$, where $\gamma$ is the relevant PB growth rate, a negligible reflectivity reduction is predicted.
As $\Omega$ increases and becomes comparable to $\gamma$, the polarization rotates within a few e-foldings, effectively reducing the resonant interaction length, thereby reducing also the reflectivity.
However, for higher frequencies, say $\Omega>\gamma$, the incident laser polarization returns to be parallel to the amplified noise many times within one e-folding and, on average, the effective interaction length is constant.
Therefore, no further reduction is then expected, \ie $R$ saturates.
This dependency is illustrated in \Fig{fig3} in which $R$ is plotted versus $\Omega$. 
It is shown that the reduction begins when $\Omega \lesssim \gamma$, 
reach to a minimum at $\Omega \approx \gamma$, and saturates for $\Omega>\gamma$.
In this regime, $\gamma$ is the SBS growth rate (because SRS is Landau damped) having the value of $\gamma= 2.16 \times 10^{12}\, {\rm sec}^{-1} $.
\begin{figure}[tb]  
    \includegraphics[width=9.0cm,trim=2.5cm 5mm 2.5cm 5mm]{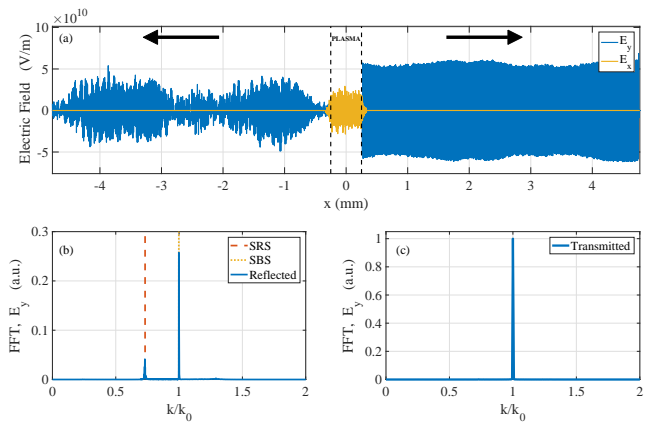}
	\caption{(color online) Constant polarization. 
	(a) Spatial distribution of the electric fields at the time the incident pulse exit the plasma. 
	(b) Spectral distribution of the reflected ($x<-0.25$ mm) electromagnetic field, $E_y$.
	(c) Spectral distribution of the transmitted ($x>0.25$ mm) electromagnetic field, $E_y$. }
	\label{fig1}
\end{figure}

\begin{figure}[tb] 
    \includegraphics[width=9.0cm,trim=2.5cm 5mm 2.5cm 5mm]{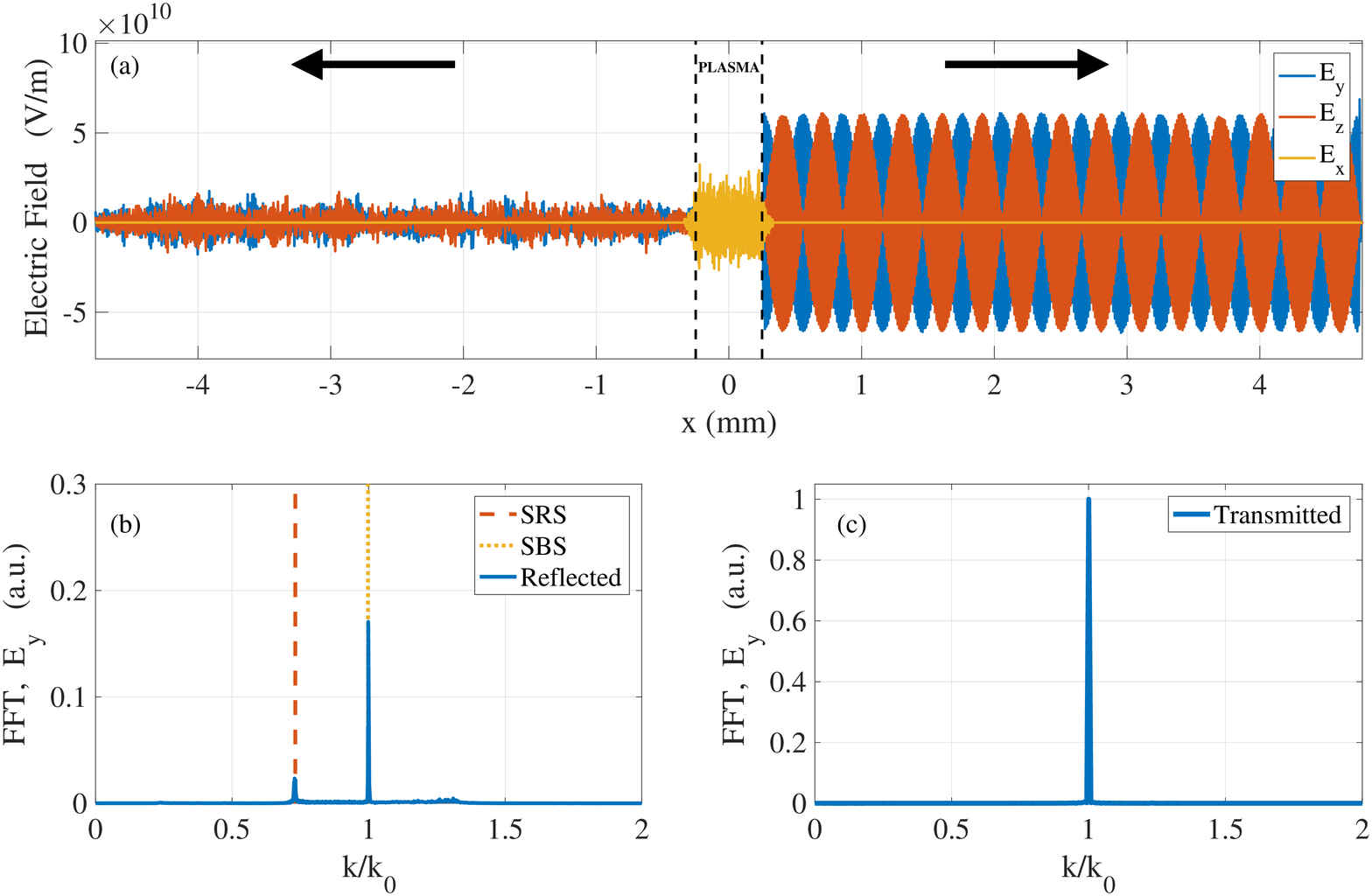}
	\caption{(color online) Rotating polarization.
	(a) Spatial distribution½ of the electric fields at the time the incident pulse exit the plasma. 
	(b) Spectral distribution of the reflected ($x<-0.25$ mm) electromagnetic field, $E_y$.
	(c) Spectral distributabletion of the transmitted ($x>0.25$ mm) electromagnetic field, $E_y$. }
	 \label{fig2}
\end{figure}

\Fig{fig3} shows another interesting (but smaller) effect in which the saturation of $R$ for $\Omega>\gamma$ is not monotonic. 
The oscillations in $R$ might be understood by introduction another time scale that may affect the reflectivity. 
Noting that the passage time of the laser front through the plasma is  $L/c$, we define the revolution frequency, $\Omega_{\rm rev}=2\pi c /L$ as the rotation frequency in which the polarization performs one revolution when crossing the plasma.
Therefore, for $\Omega=\Omega_{\rm rev}$ the incident pulse polarization performs a half revolution (\eg from $\hat{y}$ to $-\hat{y}$)  relative to a point on the counter-propagating backscattered pulse.
The phase difference, $-1=\exp(i\pi)$ is compensated after a very short time of $\pi/\omega_0\approx 0.5$fs and the parametric instability returns to be resonant.
For the example shown in \Fig{fig3}, $\Omega_{\rm rev}/2\pi=600$GHz (dotted purple line), which is when the reflectivity increases from $0.03$ to $0.05$. 

\section{Theory} \label{sec4}
In order to estimate the reduction in reflectivity, $R$, consider one of the polarizations, say $\hat{y}$, of the reflecting light, $\mathbf{A}_1$, that is amplified from the background noise. 
The idea is to find the modified linear growth rates and then to compare the amplification of the backscattered intensity with and without rotation.
It will be shown that, for rotating polarization, the effective growth rate is smaller, and thus, the total reflectivity is reduced.
For simplicity, we assume non relativistic particles and neglect collisional, kinetic, stochastic, and multidimensional effects.
However, the comparison with PIC simulations, where all of these effects (except multidimensionality and collisions) are included, provides the physical justification of using this simple model to estimate the reflectivity reduction.

\begin{figure}[tb]  
\includegraphics[width=8.5cm,trim=1.5cm 5mm 1.8cm 5mm]{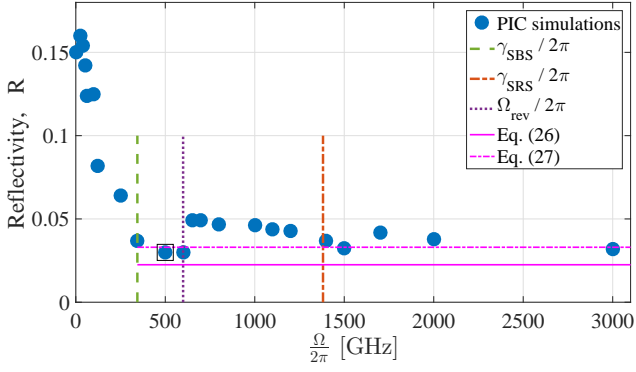}
\caption{(color online) Reflectivity versus polarization rotation frequency.
The reflectivity reduction occurs at the SBS growth rate (green dashed line) while at the SRS growth rate (red dashed-dotted line), $R$ is already saturated.  
Also shown, the theoretical predictions for the reflectivity reduction (magenta solid and dashed-dotted lines) and the revolution frequency (purple dotted line).
The black square represents the example of \Fig{fig2}.}
\label{fig3}
\end{figure}

We employ a 1D fluid model to describe the electron (ion) density perturbations, $n_e$ ($n_i$) normalized by the unperturbed density, $n_0$. 
The Maxwell wave equation describes the reflected light (dimensionless) vector potential, $\mathbf{A}_1$, in an unmagnetized plasma while the incident light, $\mathbf{A}_0$, is considered constant in the forthcoming linear analysis \cite{Forslund,Kruer}. 
\begin{eqnarray} 
  \left(\partial_t^2+\omega_e^2-c^2\partial_x^2\right) \mathbf{A}_1 &=& 
                    - \omega_e^2 n_e \mathbf{A}_0  \label{A1_a}  \\
  \left(\partial_t^2+\omega_e^2- \gamma_e v_e^2 \partial_x^2\right) n_e &=& \omega_e^2 n_i  +  
                       c^2   \, \partial_x^2 \left( \mathbf{A}_ 0\cdot \mathbf{A}_1\right) \label{n_e1}  \\
  \left(\partial_t^2+\omega_i^2-3v_i^2 \partial_x^2\right) n_i &=& \omega_i^2 n_e  .   \label{n_i1}
\end{eqnarray} 
Here, $\omega_e$ ($\omega_i$) is the electron (ion) plasma frequency; 
$n_0$ is the unperturbed average  plasma density; 
$v_e$ ($v_i$) is the electron (ion) thermal velocity;
and $\gamma_e=3$ ($\gamma_e=1$) for adiabatic (isothermal) equation of state associated with EPW (IAW).
By using the standard harmonic approximation for the electron and ion density perturbations,
\begin{equation}
   n_{e,i} = \Re \left[ \tilde{n}_{e,i} e^{ i(\omega t - k x)} \right],
\end{equation}
and for the vector potential of the reflected light
\begin{equation} \label{A1}
   \mathbf{A}_1 =   \Re \left[ a_1 \,\hat{y} \,  e^{i(\omega_1 t+k_1 x)} \right],
\end{equation}
and by neglecting the interaction terms in the right hand sides of Eqs.~(\ref{n_e1}) and (\ref{A1_a}), we get the dispersion relations for the two types of longitudinal plasma waves and for the electromagnetic (EM) waves in plasmas
\begin{eqnarray} 
  \omega^2 &=& \omega_e^2+3v_e^2k^2 \quad\quad {\rm (EPW)}   \label{EPW}  \\
  \omega    &=& c_s k  \quad\quad\quad\quad\quad \; \, {\rm (IAW)}  \label{IAW}  \\
  \omega_{0,1}^2 &=& \omega_e^2+c^2 k_{0,1}^2 \quad\quad {\rm (EM)} \label{EM}.
\end{eqnarray} 
For EPW we have neglected the ions motion while for IAW we have assumed $\omega\ll\omega_e^2$, neglected the ion thermal velocity, $v_i$, and defined $c_s=\sqrt{T_e/m_i}$. 

Next, we calculate the linear growth rate for rotating polarization incident pulse.
Note that because the incident light, $\mathbf{A}_0$, in Eq.~(\ref{A0}) is (locally) linearly polarized, we considered linear ($\hat{y}$) polarization also for the backscatter wave in \Eq{A1}. 
This polarization is not time dependent though.
Therefore, a factor of $\cos \Theta $ from \Eq{p} will appear in the right hand side of Eqs.~(\ref{A1_a}) and (\ref{n_e1}).
By assuming perfect backward resonance conditions,
\begin{gather} 
   \omega_1=\omega_0-\omega \\   
   k_1=k-k_0,
\end{gather} 
Eqs.~(\ref{A1_a})-(\ref{n_i1}) become
\begin{eqnarray} 
  \left( \omega_1^2-\omega_e^2 - c^2 k_1^2\right) a_1 &=& 
                     \frac{\omega_e^2}{2} n_e a_0 \cos\Theta \label{A1_b}  \\
  \left(\omega^2-\omega_e^2 - \gamma_e v_e^2 k^2\right) n_e + \omega_e^2 n_i &=&    -  \frac{ c^2 k^2 }{2 }  a_0 a_1 \cos\Theta  \label{n_e_b}  \\
  \left(\omega^2-\omega_i^2 - 3v_i^2 k^2\right) n_i + \omega_i^2 n_e&=& 0 .   \label{n_i_b}
\end{eqnarray} 
By solving Eqs.~(\ref{A1_b})-(\ref{n_i_b}) for $\omega$ one finds the linear growth rate, $\gamma ={\rm Im}(\omega)$. 
For SRS the ion motion is neglected, \ie $n_i=0$, and the growth rate is \cite{Forslund,Kruer}. 
\begin{gather}  \label{SRS}
    \gamma = \frac{a_0 \, c \, k \, \sqrt{\omega_e}} {4 \sqrt{\omega_b}}  \, |\cos \Theta |, \quad\quad (\rm SRS)
\end{gather} 
where we neglected thermal corrections and used the  EPW (\Eq{EPW})  and the EM (\Eq{EM}) dispersion relations. 
Similarly, for SBS one finds \cite{Forslund,Kruer}
\begin{gather} \label{SBS}
    \gamma = \frac{a_0 \, c \, k \, \omega_i} {4 \sqrt{\omega_s\omega_0}}  \, |\cos \Theta |,  \quad\quad (\rm SBS)
\end{gather} 
where $\omega_s=k c_s$ and the dispersion relations of IAW (\Eq{IAW}) and EM waves (\Eq{EM}) are used.

If the frequency is sufficiently slow, \ie $\Omega \ll \omega_0$, we can regard $\gamma$ as a weakly time dependent growth rate. 
This timescale separation allows us to define an effective e-folding
\begin{gather} \label{Gamma1}
  \Gamma=2 \int_0^{\frac{L}{c}} \gamma dt.
\end{gather}
The noise intensity is then amplified by a factor of $e^\Gamma$.
In the limit of constant polarization $\Theta=0$, we get the usual e-folding definition,
\begin{gather} \label{Gamma2}
  \Gamma_0=2 \gamma_0 L/c.
\end{gather}
Here, $\gamma_0=\gamma(\Theta=0)$ and we assumed that the noise is being amplified through the whole plasma length.
In reality, many noise sources throughout the plasma contribute to the reflectivity.
Therefore, we replace $\Gamma_0$ in \Eq{Gamma2}  by an effective e-folding, $\bar{\Gamma}_0<\Gamma_0$. 
For homogeneous noise source, this correction can be calculated by defining the average gain,
\begin{gather} \label{e_Gamma_bar}
   e^{\bar{\Gamma}_0}=\frac{ 1 }{L}\int_0^Le^{2\gamma_0 x /L} dx = \Gamma_0^{-1}\left(e^{\Gamma_0}-1\right).
\end{gather}
For $\Gamma_0>3$, this can be simplified into  
\begin{gather} \label{Gamma_bar}
   \bar{\Gamma}_0\approx\Gamma_0-\ln\Gamma_0. 
\end{gather}
By including the correction of \Eq{Gamma_bar}, assuming many polarization rotations in plasma,  $\Omega\gg2c/L$,   \Eq{Gamma1} becomes
\begin{gather} \label{Gamma3}
    \Gamma=\frac{2}{\pi} \bar{\Gamma}_0.
\end{gather}
As a result, the total amplification of the backscatter noise is effectively reduced. 
In the linear regime, the backscatter intensity can be written as $I_{\rm reflect} = I_{\rm noise} e^\Gamma$, where $I_{\rm noise}$ is the initial (thermal) noise intensity.
We note that at a given time and position, only one component of the noise polarization can be amplified, and therefore the total reduction factor is the same as that of the $\hat{y}$ polarization in \Eq{Gamma3}.
Therefore, the reduced  reflectivity can be estimated as
\begin{gather} \label{Reflectivity}
    R=e^{\Gamma - \bar{\Gamma}_0} R_0 \approx e^{-0.36 \bar{\Gamma}_0} R_0,
\end{gather}
where, $R_0$ is the reflectivity of the same incident pulse but without rotating the polarization.

Importantly, \Eq{Reflectivity} means that the reflectivity reduction factor, $R/R_0$, decreases exponentially with $\Gamma_0$.
In other words, for systems with smaller noise level but with more e-foldings, the reduction effect is more significant.
However, since we have linearized the fluid model by neglecting the nonlinear terms in Eqs.~(\ref{A1_a})-(\ref{n_i1}),
our estimation for the reflectivity reduction, \Eq{Reflectivity}, is valid only in the linear regime, $R_0\ll1$.

Finally, we estimate the reduction factor for the example presented in  Fig.~{\ref{fig2}. 
For this case, SRS is Landau damped due to high electron temperature and thus, the relevant PB is SBS.
Without rotation, the SBS growth rate for this example is $\gamma_0=2.16 \times 10^{12}\, {\rm sec}^{-1} $ resulting in
 $\bar{\Gamma}_0 = 5.2$ e-foldings. 
Therefore, we estimate the reflectivity reduction as $R/R_0 = 0.15$. 
For reflectivity without rotation (found in PIC simulations), $R_0=0.15$, the theoretical prediction for the reflectivity in the presence of polarization rotation is $R=0.023$.
As shown in \Fig{fig3} (magenta doted line) this estimation fairly agrees with the reflectivity obtained in PIC simulations, $R(\Omega/ 2 \pi = 500 {\rm GHz})=0.03$, which is  associated with reflectivity reduction of $R/R_0=0.2$.

\section{Spectral approach}   \label{sec5}

Let us discuss an alternative point of view. 
The theory that was introduced in \Sec{sec4} focuses on calculating the effective e-folding that is defined in \Eq{Gamma1}.
However, Eqs.~({\ref{A_pm})-(\ref{A_pm_a}) imply that every rotation polarization beam is a sum of two counter-rotating circularly polarized beams with a frequency shift, regardless the technique this beam is engineered.
This gives rise to an alternative, spectral-based, point of view on the reflectivity reduction effect.
In this approach, we look on the rotationally polarized beam as a sum  two, noninteracting, counter-rotating circularly polarized waves.
We note that for sufficiently large frequency shift, the two beams cannot drive the same plasma wave.
Therefore a superposition of two plasma waves with different frequencies is required for backscattering of the two incident beams.
Each interaction can be described by a different set of three-waves interaction.
Since the intensity of each frequency is half the total intensity, the dimensionless amplitude, $a_0$, is smaller by a factor of $\sqrt{2}$.
According to \Eq{SRS} for SRS and \Eq{SBS} for SBS the growth rate, $\gamma$, is reduced by the same factor.
Employing similar reasoning of the previous section for circular polarization beams results in the replacement of the $2/\pi \approx 0.64$ factor in \Eq{Gamma3} by $1/\sqrt{2} \approx 0.71$.
As a result, the spectral approach estimation for the reflectivity reduction is
\begin{gather} \label{Reflectivity_spectral}
    \frac{R}{R_0} \approx e^{-0.29 \bar{\Gamma}_0}.
\end{gather}
For the example presented in \Fig{fig2}, $\bar{\Gamma}_0=5.2$ , thus the predicted reduction is $R/R_0=0.22$, which is higher than the temporal-based estimation in \Eq{Reflectivity} and agrees better with the PIC simulations in this example. Both estimations are depicted in \Fig{fig3}.

\section{Additional results and discussion} \label{sec6}

In the previous sections we focused on the rotation polarization in a regime where the SBS is dominant.
In this section, we briefly extend the discussion to other regimes.

The first regime we consider is when the electron temperature is lower such that SRS is not Landau damped and therefore, dominant over SBS. 
To this end, we reduced the electron temperature in the PIC simulations to $T_e=100$ eV and used the following parameters: 
 $\lambda_0=0.351 \, \mu$m, $I_0=10^{15}$~W/cm$^2$, $\tau=3$~ps, $L=0.5$~mm, $n_0=2\times10^{20} \,{\rm cm}^{-3}$, and $T_i=50$ eV.
For this example, the reflectivity for fixed linear polarization was found to be $R=0.28$.
Fourier analysis of the reflected beam reveals that about $90\%$ of reflected power is due to SRS while SBS is responsible for only $10\%$.
By rotating the  polarization with frequency $\Omega/2\pi = 1600$GHz, which is larger than the SRS growth rate [\Eq{SRS}]  without rotation, $\gamma_{\rm SRS}=  2\pi \times 1560 GHz$,  the numerical reflectivity reduces to $R=0.1$.
Notable, both SRS and SBS are reduced by about factor of three.
This is a significant reduction and similar to the results observed in the SBS regime presented in \Sec{sec3}. 
However, the theoretical estimation, derived in \Sec{sec4} is not applicable for this example because the incident beam is partially depleted during the backscattering and therefore, we cannot assume a constant pump to find the linear response.

Second, for rotating, but locally linear, polarization we considered above $\Omega\ll\omega_0$.
However, higher values of $\Omega$, that are not included in our model, might be of interest. 
For example, when $\Omega>\omega_0-\omega_e$ the lower frequency, $\omega_{-}=\omega_0-\Omega$, becomes smaller than the plasma frequency, $\omega_e$, and therefore can not penetrate the plasma.
Moreover, in the limit $\Omega=\omega_0$, which is not included in our model, we formally get $\omega_{-}=0$.
In this case, the beam that is defined in \Eq{A_pm_a} becomes circularly polarized with a single frequency  $\omega_{+}=2\omega_0$ that carries all the energy.
For completeness of the discussion, we numerically compare linearly polarized and circularly polarized beams with the same wavelength.
PIC simulation shows that the reflectivity for the example with the same physical and numerical parameters as in \Fig{fig1} but with circular polarization is $R=0.2$.
This value is comparable to the reflectivity of the linear polarization, $R=0.15$, as predicted by the three-wave interaction fluid model.
Additionally, it is notable that reducing the laser wavelength reduces also the reflectivity for both linear and circular polarizations. 

The third regime is the beat wave.
Instead of rotation polarization beam we consider a beat wave, \ie a sum of two frequency-shifted linearly polarized beams (see \Sec{sec2}).
Additional PIC simulations, for the beat wave with the same parameters of the rotation polarization example of \Fig{fig2}, show a reduction of the total reflectivity by a factor of $3$  to $R=0.05$, where the total incident pulse energy is kept constant. 
This smaller reduction might be because that for the same average power the peak intensity is higher, so the growth rate is higher as well. 
However, a future study is required to optimize the pulse shape for minimum reflectivity including both time dependent envelope and polarization that opens a new dimension for such optimization.
Remarkably, besides the total reflectivity reduction that is subjected to future optimization, the rotating polarization is advantageous over the linearly polarized beat wave because it has a constant intensity.
The absence of beating at constant power density would limit deleterious effects, particularly those arising from high order effects in the wave amplitude.

Finally, we note that another possible type of beam is the sum of two linearly-polarized frequency shifted waves but with perpendicular polarizations. 
Similarly to the rotation polarization case, the total intensity in this case is constant but the time dependent  polarization is switching between linear and circular polarizations due to the relative phase accumulations between the two waves. 
Based on the results here, one would expect that the reflectivity in this case would also be similarly reduced.

 \section{Conclusions} \label{sec7}
In summary, we show that by rotating the polarization of an incident light on a plasma, the overall reflectivity can be significantly reduced.
For example, PIC simulations show, that for parameters similar to those of an indirect ICF, polarization rotation can reduce the SBS reflectivity by a factor of $5$.
An analytical estimation, based on the linearized fluid model, agrees with the numerical simulations.
In order to get significant reflectivity reduction, the rotation frequency must be of the order of the dominant instability growth rate, but the reduction saturates for higher frequencies.
Our theory predicts that the reduction effect is more significant for systems with larger number of e-foldings, provided the system is in linear regime, \ie $R_0\ll1$.
Such polarization rotation can be realized by combining two frequency shifted, counterrotating circularly polarized lasers.
Also, in addition to amplitude modulations and polarization rotation, different polarization modulations (\eg oscillations between linear and circular polarizations) of the incident light are predicted to result in a similar reflectivity reduction while a future study is required to optimized the best method.
Notably, in all of these techniques, the modulated beam comprises two frequency shifted waves but with different polarization.
Therefore, in identifying the source of the backscatter reduction, it is difficult to disentangle the polarization or amplitude modulation from the frequency separation based approach used to engineer the beam.
One advantage of polarization rotation over other techniques is its constant intensity despite the frequency separation.
In general, the polarization rotation technique opens a new dimension of modulating the incident pulse for reducing parametric reflectivity, which is important for ICF experiments or for other systems where it is essential to reduce backscattering of light passing through a plasma or, for that matter, other dielectric media.

\acknowledgments{
The authors appreciate discussions with P. Michel.
This work was supported by 
NNSA Grant No.~DE-NA0002948, 
AFOSR Grant No.~FA9550-15-1-0391, 
and DOE Contract No.~DE-AC02-09CH11466.
}


\end{document}